\documentclass[a4paper,11pt]{article}
\usepackage{jheppub}
\usepackage{eurosym}
\usepackage{amssymb}
\usepackage{amsfonts,cite}
\usepackage{amsmath}
\usepackage{graphicx}
\usepackage{multirow}
\usepackage{xcolor}
\usepackage{epsfig}
\usepackage{fancyhdr}

\newbox\mybox

\newcommand\fverb{\setbox\mybox=\hbox\bgroup\verb}
\newcommand\fverbdo{\egroup\medskip\noindent\fbox{\unhbox\mybox}\ }
\newcommand\fverbit{\egroup\item[\fbox{\unhbox\mybox}]}

\abstract{We investigate an interacting Pais-Uhlenbeck oscillator with a Landau-Ginzburg type
	interaction term and analyse its classical dynamics from a geometric and numerical point of view.
	We show that the resulting fourth-order equation of motion admits a conformal
	bi-Hamiltonian formulation, possesses a non-trivial set of Lie symmetries and we demonstrate the existence of bounded and regular trajectories in representative parameter regimes.
	By establishing an explicit correspondence with an integrable generalised
	H\'enon-Heiles system, we show that the interacting higher–derivative dynamics
	inherits the integrability properties of the latter.
	This connection allows us to construct a second conserved Hamiltonian, to clarify the
	geometric origin of separability, and to obtain explicit classical solutions in terms
	of elliptic functions. Our results provide a concrete example of an interacting higher-derivative system for which integrability and periodic classical solutions can be established in a fully
	explicit manner.}    
	
\author[a]{Alexander Felski}
\author[b]{and   Andreas Fring}

\emailAdd{alexander.felski@mpl.mpg.de}
\emailAdd{a.fring@city.ac.uk}	

\title{Conformal bi-Hamiltonian structure and integrability of an interacting Pais-Uhlenbeck oscillator}

\affiliation[a]{Max Planck Institute for the Science of Light, Staudtstraße 2,91058 Erlangen, 	Germany}
\affiliation[b]{Department of Mathematics, City St George's, University of London,  Northampton Square, \\ London EC1V 0HB, UK}

\begin{document}
		\maketitle
	
	\pagestyle{fancy}
	\fancyhead{} 
	\fancyhead[LE,RO]{\small\itshape  Conformal bi-Hamiltonian structure and integrability of an interacting Pais-Uhlenbeck oscillator} 
	
	\renewcommand{\headrulewidth}{0.4pt}
	
\section{Introduction}	

The Pais-Uhlenbeck (PU) oscillator \cite{pais1950field}  constitutes a prototypical example of a dynamical system with higher time derivatives and has long served as a testing ground for
conceptual and technical issues in higher-derivative dynamics.  Owing to the presence
of derivatives of order higher than one in the Lagrangian, the model naturally
exhibits the characteristic features of Ostrogradsky-type systems, and provides a
simple laboratory in which questions concerning stability \cite{ghostconst,motohashi1,motohashi4}, Hamiltonian formulation \cite{bolonek2005ham,bender2008no,smilga2009comments,most2010h,mosta2011im,andrzejewski2014ham,Sugg3,smilga2017class}
and quantisation \cite{fring2025quant} can be studied in a controlled setting.

For the non-degenerate free PU model, corresponding to distinct
frequencies, the classical and quantum theory is by now fairly well understood. In
particular, it has been shown that the system can be mapped, by a suitable canonical \cite{FFT,felski2026three}
and similarity transformation \cite{fring2025ghost}, to a Hamiltonian formulation which is bounded from
below and admits a consistent probabilistic interpretation. These results
demonstrate that the presence of higher time derivatives does not by itself preclude
a physically meaningful description.

The situation becomes substantially more subtle in the degenerate case, where the
frequencies coincide. In this regime the two normalisable sectors of the theory disappear. The spectral and dynamical properties change
qualitatively and the construction of a well-defined Hamiltonian framework is more
challenging. Recent investigations have shown that even in the degenerate case a
consistent spectral description may be achieved, although the underlying structure is
considerably more intricate than in the non-degenerate setting
\cite{fring2026spectrum}.

The real challenge, however, arises once interactions are introduced. It is well
known that generic interaction terms tend to destabilise higher-derivative systems
and typically lead to runaway solutions or to the rapid amplification of the ghost
modes associated with the Ostrogradsky construction
\cite{Sugg1,pavvsivc2013pais,pavvsivc2016pais}. As a consequence, interacting
PU models are usually regarded as dynamically problematic and only very
few examples are known in which nontrivial interactions can be incorporated without
destroying all analytic control. 

In the present work we investigate a class of interacting PU models with
a Landau-Ginzburg type interaction term for which the dynamics retains a remarkably
rich geometric structure. In particular, we show that the interacting system admits a
conformal bi-Hamiltonian formulation and can be related, through an explicit
elimination procedure, to an integrable generalised H\'enon-Heiles system. This
connection allows us to transfer powerful geometric tools from finite-dimensional
integrable systems, such as conformal bi-Hamiltonian structures, Lie symmetries and
separation of variables, to the interacting higher-derivative PU
dynamics. As a result, we obtain an explicit integrable realisation of an interacting
higher-derivative model and construct families of periodic classical solutions for
the corresponding fourth-order equations of motion. We also perform a direct numerical analysis of the fourth-order PU equation, which confirms the existence of bounded and regular trajectories in representative parameter regimes.

Our manuscript is organised as follows. In section 2 we introduce the interacting PU model with a Landau-Ginzburg type interaction and derive its Hamiltonian formulation. We show that the resulting higher-derivative dynamics admits a conformal bi-Hamiltonian representation and determine its Lie point symmetries. In section 3 we review the integrable generalised H\'enon-Heiles system and establish an explicit correspondence between this model and the interacting PU equation, together with the associated conformal bi-Hamiltonian and symmetry structures. In section 4 we investigate the classical dynamics of the interacting PU equation. We first present a direct numerical analysis of the fourth-order equation, illustrating the existence of bounded and regular trajectories for representative parameter regimes. We then construct the classical solutions analytically by exploiting the integrability inherited from the H\'enon–Heiles system and by performing an explicit separation of variables. Finally, section 5 contains our conclusions and an outlook.
	
\section{The Pais-Uhlenbeck model with Landau-Ginzburg interaction term}	

We consider a higher time derivative PU oscillator supplemented by a polynomial Landau-Ginzburg type self-interaction. We introduce the Lagrangian
\begin{equation}
	{\cal L}_{\text{PUI}}(q,\dot{q} , \ddot{q})= \frac{\ddot{q}^2}{2}-\frac{1}{2} {\alpha} \dot{q}^2+ \frac{1}{2} {\beta}  q^2+\frac{1}{3} g \left[ 5 q \dot{q}^2-2 E q- {\alpha} q^3+\frac{5
		}{6} g q^4   \right],  \label{HHinter}
\end{equation}	
where $\alpha, \beta$ are the standard PU parameters, $g$ controls the interaction strength and $E$ is a constant which will later be identified with an energy parameter arising from the reduction of the H\'enon-Heiles system. The interaction is of Landau–Ginzburg type in the sense that it consists of polynomial self-couplings of the coordinate $q$, supplemented by a derivative-dependent interaction term which does not introduce derivatives beyond those already present in the free PU model.  For $g \rightarrow 0$ all interaction terms disappear and we recover the standard quadratic PU Lagrangian with frequencies determined by $\alpha, \beta$. 

Since the Lagrangian depends on second time derivatives, the corresponding equation of motion is obtained from the generalised Euler–Lagrange equation
\begin{equation}
	\frac{\partial \cal{L}}{\partial q}
	-\frac{d}{dt}\frac{\partial \cal{L}}{\partial \dot{q}}
	+\frac{d^{2}}{dt^{2}}\frac{\partial \cal{L}}{\partial \ddot{q}}
	=0 ,
\end{equation}
which for $	{\cal L}_{\text{PUI}}(q,\dot{q} , \ddot{q})$ yields the fourth-order ordinary differential equation
\begin{equation}
	\ddddot{q} +{\alpha} \ddot{q}+{\beta}   q-\frac{10}{3}  g q \ddot{q}-\frac{5 g \dot{q}^2}{3}-\frac{2 E g}{3}-g {\alpha}  q^2+\frac{10}{9} 
	g^2  q^3 =0 . \label{equmPUI}
\end{equation}

Following Ostrogradsky's approach \cite{ostrogradsky1850memoire} we carry out a Legendre transformation and convert the second-order Lagrangian (\ref{HHinter}) into a first-order Hamiltonian system with two configuration variables and a four-dimensional phase space. Introducing the coordinates $q_1 = q$,  $q_2 = \dot{q}$, the conjugate momenta are defined in the standard way by treating $q$ and $\dot{q}$ as independent variables
\begin{equation}
	      \pi_1 = \frac{\partial {\cal L} }{\partial \dot{q}} - \frac{d}{dt}  \frac{\partial {\cal L}}{\partial \ddot{q}}  = \frac{10}{3} g q \dot{q} - \dddot{q} - {\alpha} \dot{q},
	  \quad
	\pi_2 = \frac{\partial {\cal L}}{\partial \ddot{q}}  = \ddot{q} .   \label{equm3}
\end{equation}	

The Hamiltonian in terms of the canonical variables is obtained by the Ostrogradsky Legendre transformation 
\begin{equation}
{\cal H}(q_1,q_2,\pi_1,\pi_2;\tilde q)
= \pi_1 \dot q_1 + \pi_2 \tilde q
- {\cal L}(q_1,q_2,\tilde q),
\end{equation}
where $\tilde{q}$ denotes the second derivative $\ddot{q}$. Since $\pi_2 = \partial  {\cal L}/ \partial \ddot{q}$, the relation between $\pi_2$ and $\tilde q$ can be inverted and yields
$\tilde q=\tilde{f}(q_1,q_2,\pi_2)$. Thus, with ${\cal L}_{\text{PUI}}(q,\dot{q} , \ddot{q})$
we obtain the Hamiltonian of the two-dimensional system
\begin{equation}
	{\cal H}_{\text{PUI}}(q_1,q_2,\pi_1,\pi_2)= \pi_1 q_2+ \frac{\pi_2^2}{2}+\frac{{\alpha}}{2}   q_2^2 -\frac{{\beta}}{2}  q_1^2+ \frac{g}{3} \left[
	2 E q_1+ {\alpha}  q_1^3-5q_1 q_2^2 -\frac{5}{6} g
	q_1^4 \right] .
\end{equation}
For $g=0$ this Hamiltonian reduces to the standard quadratic PU Hamiltonian written in terms of the Ostrogradsky variables. Converting back to the original higher derivative $q$-variables this gives
\begin{equation}
	{\cal H}_{\text{PUI}}(q,\dot{q} , \ddot{q},\dddot{q})= \frac{\ddot{q}^2}{2}-\frac{{\alpha} }{2}  \dot{q}^2-\frac{{\beta}}{2} q^2 -\dot{q} \dddot{q} +\frac{g}{3}  \left[2 E q+ {\alpha} q^3 -\frac{5 }{6}g q^4 +5 q \dot{q}^2   \right] .
\end{equation}
As a consistency check, we verify that the Hamiltonian is conserved along the solutions of the fourth-order equation of motion (\ref{equmPUI}). The total time derivative reads
\begin{equation}
\frac{{d\cal H}_{\text{PUI}}}{dt}= \frac{\partial {\cal H}_{\text{PUI}}}{\partial q} \dot{q}+ \frac{\partial {\cal H}_{\text{PUI}}}{\partial  \dot{q}} \ddot{q} + \frac{\partial {\cal H}_{\text{PUI}}}{\partial  \ddot{q}} \dddot{q}+ \frac{\partial {\cal H}_{\text{PUI}}}{\partial  \dddot{q}} \ddddot{q} =0.
\end{equation}
This Hamiltonian formulation provides the starting point for the construction of the Poisson structures and the conformal bi-Hamiltonian representation discussed in the next subsection.

\subsection{Conformal bi-Hamiltonian structure}

In \cite{FFT} the use of the bi-Hamiltonian structure of the standard PU-model was extremely useful,  in particular it was utilised to identify a positive-definite Hamiltonian that
generates the system’s flow in the degenerate model. In general a dynamical system is called bi-Hamiltonian if its vector field admits two distinct
Hamiltonian representations with respect to two compatible Poisson structures.
In many integrable systems this property is responsible for the existence of recursion
operators, hierarchies of commuting flows and additional conserved quantities.
In the present interacting higher-derivative setting, we will, however, encounter a
slightly weaker structure: the same vector field can be represented as a Hamiltonian
flow with respect to a second Poisson tensor only up to a non–constant scalar factor.
Such systems are usually referred to as conformal or quasi bi-Hamiltonian systems
\cite{morosi1997quasi,crampin2002bi,falqui2003separation}.

Here, the dynamical equations of the PU oscillator with interaction term can be cast into the form
\begin{equation}
	\frac{d\vec{q}}{dt} = \vec{V}(\vec{q}),
	\label{flow0}
\end{equation}
where we introduced the phase-space vector
$
\vec{q} = ( q , \dot{q},\ddot{q}, \dddot{q} ) \in \mathbb{R}^4
$
and the associated first-order differential operator
\begin{equation}
	 V= \dot{q} \partial_q +  \ddot{q} \partial_{\dot{q}} 
	+  \dddot{q} \partial_{\ddot{q}} - \left( {\alpha} \ddot{q}-\frac{10}{3}  g q \ddot{q}-\frac{5 g \dot{q}^2}{3}-\frac{2 E g}{3}+{\beta}   q-g {\alpha}  q^2+\frac{10}{9} 
	g^2  q^3  \right) \partial_{ \dddot{q} } . \label{1flow}
\end{equation}
which acts as a vector field when interpreted as a derivation on $C^{\infty}(\mathbb{R}^4)$. 

Writing $V=\sum_{i=1}^4 v_i \partial_{q_i}$, the vector field can be generated from the
Hamiltonian ${\cal H}_{\text{PUI}}$ by means of a  non-canonical Poisson
structure according to
\begin{equation}
	\vec{v} = \left\{  \vec{q}, {\cal H}_{\text{PUI}}    \right\}       = J \cdot \nabla {\cal H}_{\text{PUI}},    \label{flow1}
\end{equation}
for the Poisson tensor
\begin{equation}
	J= \left(
	\begin{array}{cccc}
		0 & 0 & 0 & -1 \\
		0 & 0 & 1 & 0 \\
		0 & -1 & 0 & \alpha  \\
		1 & 0 & -\alpha  & 0 \\
	\end{array}
	\right)+g \frac{10}{3}  q \left(
	\begin{array}{cccc}
		0 & 0 & 0 & 0 \\
		0 & 0 & 0 & 0 \\
		0 & 0 & 0 & -1 \\
		0 & 0 & 1 & 0 \\
	\end{array}
	\right)   . 
\end{equation}
The first term corresponds to the Poisson structure of the free PU model, while the
second term represents the deformation induced by the interaction.

As in the case of the standard PU model, it is natural to ask whether there exists
a second Hamiltonian representation which generates the same dynamical flow.
In the interacting case, however, we do not require a strict bi-Hamiltonian structure.
Instead, we look for a conformal (or quasi) bi-Hamiltonian representation of the form
\cite{falqui2003separation}
\begin{equation}
	\vec{v} = f(\vec{q}) J_2 \cdot \nabla {\cal H}_2.    \label{cflow2}
\end{equation}
It is worth stressing that the conformal factor $f(\vec q)$ can be interpreted as a
reparametrisation of the time variable.
Indeed, introducing a new evolution parameter $\tau$ defined by
\begin{equation}
	d\tau = f(\vec q)\, dt ,
\end{equation}
the conformal Hamiltonian flow (\ref{cflow2}) becomes an ordinary Hamiltonian flow
with respect to $J_2$,
\begin{equation}
	\frac{d\vec q}{d\tau}= J_2 \cdot \nabla {\cal H}_2 .
\end{equation}
Therefore, the two Hamiltonian representations generate the same trajectories in
phase space, differing only by a reparametrisation of the time variable.
This property motivates the terminology \emph{conformal} bi-Hamiltonian structure.

Indeed, we find that the vector field generated by ${\cal H}_{\text{PUI}}$ also admits
a conformal Hamiltonian representation with Hamiltonian ${\cal H}_2$ given by
\begin{eqnarray}
	{\cal H}_2(q,\dot{q} , \ddot{q},\dddot{q})&=& \frac{\beta  \dot{q}^2}{2}-\frac{\alpha  \ddot{q}^2}{2}-\frac{\dddot{q}^2}{2}-q \beta  \ddot{q} +g \left[\frac{2 q^3 \beta }{3}+\frac{2 E \ddot{q}}{3}+q^2 \alpha  \ddot{q}-\frac{1}{3} \dot{q}^2 \ddot{q}+\frac{2 q \ddot{q}^2}{3}+2 q \dot{q} \dddot{q} \right] \notag \\
	&& + g^2 \left[\frac{2E q^2}{3}+\frac{q^4 \alpha }{2}+\frac{5}{3} q^2 \dot{q}^2+\frac{10 q^3 \ddot{q}}{9} \right] + g^3 \frac{4 }{9}  q^5,
\end{eqnarray}
with Poisson tensor
\begin{equation}
	J_2 = \left(
	\begin{array}{cccc}
		0 & \frac{1}{\beta } & 0 & 0 \\
		-\frac{1}{\beta } & 0 & 0 & 0 \\
		0 & 0 & 0 & -1 \\
		0 & 0 & 1 & 0 \\
	\end{array}
	\right)+ \frac{2 g}{\beta}   \left(
	\begin{array}{cccc}
		0 & 0 & 0 & q \\
		0 & 0 & -q & -\dot{q} \\
		0 & q & 0 & \frac{1}{3} \left(5 g q^2+\ddot{q}\right) \\
		-q &\dot{q}& \frac{1}{3} \left(-5 g q^2-\ddot{q} \right) & 0 \\
	\end{array}
	\right),
\end{equation}
and conformal factor
\begin{equation}
    f(\vec{q}) = 	\frac{3 \beta }{3 \beta +2 g^2 q^2-2 g  \ddot{q}} . \label{confact}
\end{equation}
The tensor $J_2$ is field-dependent and contains interaction-induced
terms of order $g$ and higher. The non-trivial dependence of $f(\vec q)$ on the phase-space variables is the origin of
the conformal character of the bi-Hamiltonian structure.

When taking the limit $g\rightarrow 0$, the conformal factor tends to unity and we fully
recover the genuine bi-Hamiltonian structure of the free PU model discussed in
\cite{FFT}. 

Furthermore, we verify that both tensors $J$ and $J_2$ satisfy the Jacobi identity
\begin{equation}
 J^{i\ell}\,\partial_\ell J^{jk}
+ J^{j\ell}\,\partial_\ell J^{ki}
+ J^{k\ell}\,\partial_\ell J^{ij} =0, \qquad i,j,k \in \{1,2,3,4  \}, \, i\neq j, \, j \neq k, \,\, i \neq k ,
\label{Jacobiid}
\end{equation}
and are therefore genuine Poisson tensors.

As an additional consistency check we consider Jacobi’s formula for the time derivative
of the determinant of a Poisson tensor. When $J$ is invertible it reads
\begin{equation}
    \frac{d}{dt} \left( \det J   \right)  = \det J  \,\text{tr} \left( J^{-1}   \frac{dJ}{dt}     \right),
\end{equation}
where the derivative is understood as the Lie derivative
\begin{equation}
	\left( \frac{dJ}{dt} \right)^{ij} = \left( {\cal L}_{X_{\cal H}} J    \right)^{ij}= \partial_k {\cal H} \left( J^{i\ell}\,\partial_\ell J^{jk}
	+ J^{j\ell}\,\partial_\ell J^{ki}
	+ J^{k\ell}\,\partial_\ell J^{ij} \right).
\end{equation}
Here ${\cal L}_{X_{\cal H}}$ denotes the Lie derivative along the Hamiltonian vector
field $X_{\cal H}=J\cdot\nabla{\cal H}$.

Thus, whenever the Jacobi identity is satisfied, the determinant of the corresponding
Poisson tensor is preserved along the Hamiltonian flow.
For the tensor $J$ this is trivially realised, since $ \det J=(\mathrm{Pf} J)^2=1 $ is constant.
In contrast, in the conformal bi-Hamiltonian setting this property is lost. Indeed, for the second Poisson tensor we obtain $ \det J_2 =  (\mathrm{Pf} J_2)^2=1/(\beta f)^2$, which is no longer a constant of motion.

\subsection{Lie symmetries of the vector flow}

We now determine the Lie point symmetries of the first-order dynamical system
(\ref{flow0}). Such symmetries are generated by vector fields on phase space whose
flow maps solutions of the dynamical system into solutions.
Equivalently, a Lie symmetry is characterised by the invariance of the dynamical
vector field under the corresponding infinitesimal transformation.

We identify the Lie symmetries in the usual way.
Infinitesimally transforming $q_i \rightarrow \tilde{q}_i = q_i + \varepsilon \xi_i$,
with $\varepsilon \ll 1$, the component form of (\ref{flow0}) transforms as
\begin{equation}
	\frac{d \tilde{q}_i}{dt} = V_i(\tilde{q}) - \delta V_i(\tilde{q}), \qquad \text{with} \,\,\,\,  \delta V_i(\tilde{q}) = \varepsilon \left[X,V\right]_i(\tilde{q}) = \varepsilon  \sum_{j=1}^4   \left(  \xi_j \frac{\partial v_i}{ \partial q_j} - v_j \frac{\partial \xi_i}{ \partial q_j} \right),
\end{equation}
where $[X,V]$ denotes the Lie bracket of the vector fields $X$ and $V$ on
$\mathbb{R}^4$ and the Lie algebraic generators are defined as
\begin{equation}
	X = \sum_{i=1}^4 \xi_i \partial_{q_i}  .
\end{equation}
The functions $\xi_i(\vec q)$ are determined by imposing the invariance condition
$[X,V]=0$. Using a polynomial Ansatz for the components $\xi_i(\vec q)$, containing all monomials
up to cubic order and selected quartic terms in the phase-space coordinates $q_i$,
we solve the determining equations following from $[X,V]=0$. In this way we obtain two linearly independent solutions, one of which corresponds
to the dynamical vector field itself, and a second genuinely non-trivial symmetry
\begin{eqnarray} 
	X_1 &=& \dot{q} \partial_q+\ddot{q} \partial_{\dot{q}}
	+\dddot{q} \partial_{\ddot{q}}+\left(\frac{2 E g}{3}-\frac{10 g^2
		q^3}{9}+g q^2 \alpha -q \beta +\frac{5 g \dot{q}^2}{3}+\frac{10}{3} g q \ddot{q}-\alpha  \ddot{q}\right)    \partial_{\dddot{q} }, \\
	X_2 &=& \left(\dddot{q}-2 g q \dot{q}+\alpha  \dot{q}\right)  \partial_q
	+\left[\frac{g}{9} \left(6 E-10 g q^3+9 q^2 \alpha -3 \dot{q}^2+12 q
	\ddot{q}\right) -q \beta \right] \partial_{\dot{q}}    \\
	&& \!\!\!	\!\!\!	 \!\!\!	 \!\!\!	 +\frac{1}{3} \left[4 g q \dddot{q}-10 g^2 q^2 \dot{q}-3 \beta  \dot{q}+2 g \dot{q} \left(3 q \alpha
	+\ddot{q}\right)\right]  \partial_{\ddot{q}}   + \frac{1}{27} \left\{    -9 \beta  \left(4 g q^2+3 \ddot{q}\right)       \right.    \notag \\
&&  \!\!\!	\!\!\!	 \!\!\!	 \!\!\!	 +\left.  2 g \left[12 E g q-20 g^2 q^4+27 \dot{q}
	\left(\dddot{q}+\alpha  \dot{q}\right)+9 \ddot{q} \left(q \alpha +\ddot{q}\right)+3 g q \left(q \left(6 q \alpha +5 \ddot{q}\right) -20 \dot{q}^2      \right)\right]  \right\} \partial_{\dddot{q} }  . \notag  
\end{eqnarray}
The generator $X_1$ coincides with the dynamical vector field $V$ itself and therefore
represents the trivial symmetry associated with time translations. The second generator $X_2$ represents a non-trivial Lie point symmetry of the interacting
PU flow. Both generators commute, $ [X_1,X_2]=0$, and, in addition, they leave both Hamiltonians invariant
\begin{equation}
[X_1,X_2]=0,	 \quad  X_1	\left(  {\cal H}_{\text{PUI}} \right) =0,  \quad X_1	\left(  {\cal H}_2 \right) =0, \quad
	X_2	\left(  {\cal H}_{\text{PUI}}\right) =0,  \quad X_2	\left(  {\cal H}_2 \right) =0.
\end{equation}
This compatibility between the Lie symmetry structure and the conformal
bi-Hamiltonian formulation provides an additional consistency check of the geometric
framework developed in the previous subsection.

\section{The generalised H\'enon-Heiles model}

The generalised H\'enon-Heiles model \cite{henonheils} is defined by the Lagrangian
\begin{equation}
	{\cal L}_{HH}(x,y,\dot{x},\dot{y} )= \frac{\dot{x}^2}{2}+\frac{\dot{y}^2}{2}-\frac{	\tilde{\alpha}  x^2}{2}-\frac{\tilde{\beta}  y^2}{2}-g_1 x y^2+\frac{g_2 x^3}{3} , \qquad              	\tilde{\alpha},\tilde{\beta},g_1,g_2   \in \mathbb{R} , \label{HHL}
\end{equation}
giving rise to a set of Euler-Lagrange equations in the form of the coupled set of second order differential equations
\begin{equation}
	\ddot{x}+\tilde{\alpha}  x-g_2 x^2+g_1 y^2=0, \qquad \text{and} \qquad  \ddot{y}+\tilde{\beta}  y+ 2 g_1 x y=0. \label{twoeqom}
\end{equation}
A Legendre transformation leads immediately to the Hamiltonian formulation
\begin{equation}
	{\cal H}_{HH}(x,y,\dot{x},\dot{y} )=\frac{\dot{x}^2}{2}+\frac{\dot{y}^2}{2}+\frac{\tilde{\alpha}  x^2}{2}+\frac{\tilde{\beta}  y^2}{2}-\frac{1}{3} g_2 x^3+g_1 x y^2.
\end{equation}
As noted by Fordy \cite{fordy1991henon} the coupled second order equations (\ref{twoeqom}) can be combined into a fourth order equation in one variable
\begin{equation}
	x^{(4)} + F(x,\dot{x},\ddot{x}) =0, \label{HHequom}
\end{equation}
where
\begin{equation}
	F(x,\dot{x},\ddot{x})={\tilde{\alpha}} \ddot{x}+2\left(4 g_1-g_2\right)x \ddot{x}-2 \left(g_1+g_2\right)\dot{x}^2+4 E g_1+ 	{\tilde{\beta}} x+ \left(6 \tilde{\alpha}  g_1-4 \tilde{\beta}  g_2\right) x^2  -\frac{20}{3} g_1 g_2 x^3. \label{FFF}
\end{equation}
This reduction is obtained by differentiating the $x$-equation in (\ref{twoeqom}) twice,
eliminating $\ddot y$ using the $y$-equation, eliminating $y^2$ using the original $x$-equation,
and finally eliminating $\dot y^{\,2}$ using the conserved energy $E={\cal H}_{HH}$.
 
  Notice that for $(\tilde{\alpha} + 4 \tilde{\beta}) = \alpha$, $ 4 \tilde{\alpha}  \tilde{\beta} = \beta$, $g_1=-g_2/6$, $g_2=g$, and upon identifying $x=q$,
  equation (\ref{HHequom}) reproduces precisely the interacting PU equation of motion
  (\ref{equmPUI}).
  
  Alternatively, we may solve the second equation in (\ref{twoeqom}) for $x$ as a function of $y,\ddot{y}$, differentiate the expression twice and in this way express the first equation entirely in terms of $y$ and its derivatives
  
  \begin{equation}
  	\frac{1}{2}g_2\,y\left(y\tilde{\beta}+\ddot{y}\right)^2
  	-2g_1^3\,y^5
  	+g_1\Big[
  	\left(y^2\tilde{\alpha}+2\dot{y}^2\right)\ddot{y}
  	-y\ddot{y}^2
  	+y^3\tilde{\alpha}\tilde{\beta}
  	-2y\dot{y}\dddot{y}
  	+y^2 y^{(4)}
  	\Big]=0.
  \end{equation}
  This procedure yields an alternative fourth-order scalar equation for $y(t)$, which we do not analyse further in the present work.
  
  \subsection{Conformal bi-Hamiltonian structure} 

It was noted in \cite{caboz1991bi} that precisely for the special relation
$g_1=-g_2/6$ the H\'enon-Heiles system admits a conformal bi-Hamiltonian structure.
Accordingly, we fix the normalisation $g_1=-1$ and $g_2=6$. Defining the phase-space vector $\vec{q}=(x,y,\dot{x},\dot{y})$, the flow that produces the two coupled equations in (\ref{twoeqom}) 
  \begin{equation}
  	\vec{v}=(\dot{x},\dot{y},  6 x^2 + y^2-\tilde{\alpha} x,2 x y -\tilde{\beta} y)  \label{HHflow}
  \end{equation}
  can be obtained equivalently from the two Hamiltonians 
  \begin{eqnarray}
  		{\cal H}_1 (x,y,\dot{x},\dot{y})&=& \frac{\dot{x}^2}{2}+\frac{\dot{y}^2}{2}+\frac{{\tilde{\alpha}} }{2}  x^2+\frac{\tilde{\beta}  }{2} y^2-2 x^3-x y^2, \\
  		{\cal H}_2 (x,y,\dot{x},\dot{y})&=& (4 \tilde{\beta} -\tilde{\alpha} ) \left(\dot{y}^2+\tilde{\beta}  y^2\right)+4 \left(x^2 y^2-x y^2 \tilde{\beta} -y \dot{x} \dot{y}+x \dot{y}^2\right)+y^4,
  \end{eqnarray}
  \begin{equation}
	 J_1 \cdot \nabla 	{\cal H}_1 =  \vec v =  f(y) J_2 \cdot \nabla 	{\cal H}_2 ,
  \end{equation}
  with associated Poisson bracket tensors
  \begin{equation}
  	J_1 = \left(
  	\begin{array}{cccc}
  		0 & 0 & 1 & 0 \\
  		0 & 0 & 0 & 1 \\
  		-1 & 0 & 0 & 0 \\
  		0 & -1 & 0 & 0 
  	\end{array}
  	\right), \qquad
  	J_2 = \left(
  	\begin{array}{cccc}
  		0 & 0 & \,\, 2\left(\frac{\tilde{\alpha}}{4}-\tilde{\beta}-x\right) \,\, & -y \\
  		0 & 0 & -y & 0 \\
  		\,\, -2\left(\frac{\tilde{\alpha}}{4}-\tilde{\beta}-x\right) \,\,& y & 0 & -\dot{y}\\
  		y & 0 & \dot{y}& 0 \\
  	\end{array}
  	\right)  ,
  \end{equation}
and $f(y) = 1/(4 y^2)$. We verify that $J_1$ and $J_2$ satisfy the Jacobi identity (\ref{Jacobiid}). Here ${\cal H}_1$ coincides with ${\cal H}_{HH}$ for the specified values of the coupling constants. Having obtained a second Hamiltonian we can
  also combine the two second order equations (\ref{twoeqom}) into a fourth order equation by following the same line of argumentation as before by replacing $\dot{y}^2$ where $K$ denotes the constant value of the second Hamiltonian ${\cal H}_2$. In this way we obtain
  \begin{eqnarray}
  && \ddddot{x}  (\tilde{\alpha} -4 \tilde{\beta} )  + 2 \left[K-6 \dot{x}^2 (\tilde{\alpha} -4 \tilde{\beta} )\right]+x^3 (56 \tilde{\alpha} -96 \tilde{\beta} )+x^2 \left(96 \tilde{\beta} ^2-6 \tilde{\alpha} ^2-8 \tilde{\alpha} 
  \tilde{\beta} \right)   \qquad \qquad \\ 
  &&+80 x^2 \ddot{x}+\left(\tilde{\alpha} ^2-16 \tilde{\beta} ^2\right) \ddot{x}+x (64 \tilde{\beta} -28 \tilde{\alpha} ) \ddot{x}+4 \tilde{\alpha}  \tilde{\beta}  x
  (\tilde{\alpha} -4 \tilde{\beta} )-10 \ddot{x}^2 -120 x^4 =0 . \notag
   \end{eqnarray}
   This yields yet another fourth-order scalar ordinary differential equation, structurally akin to the PU equation (\ref{equmPUI}), whose detailed analysis we leave for future work.

  \subsection{Lie symmetries of the H\'enon-Heiles equation}
  We determine the Lie point symmetries of the H\'enon--Heiles flow (\ref{HHflow})
  using the same infinitesimal invariance criterion as in the previous subsection,
  see also \cite{hydonsymm}.
  Using a polynomial Ansatz for the components $\xi_i$, cubic in the phase-space
  coordinates $q_i$, we obtain two linearly independent solutions of the determining
  equations
  \begin{eqnarray}
  	X_1 &=& \dot{x} \partial_x +     \left( 6 x^2 + y^2-\tilde{\alpha} x\right)\partial_{\dot{x}}  +  \dot{y} \partial_y +   \left( 2 x y -\tilde{\beta} y \right)  \partial_{\dot{y}} 
  	, \label{Lie1} \\
  	X_2 &=& 2 \left(\tilde{\beta}  \dot{x}+y \dot{y}\right)  \partial_x  + 2 \left[ x \left(2 y^2+6 x \tilde{\beta} -\tilde{\alpha}  \tilde{\beta} \right)+\dot{y}^2\right]   \partial_{\dot{x}}    \label{Lie2}     \\
  	&& + \left[2 y \dot{x}+(\tilde{\alpha} -2 \tilde{\beta} -4 x) \dot{y}\right]  \dot{y} \partial_y  +  \left[ \left(4 x^2+2 y^2-\tilde{\alpha}  \tilde{\beta} +2 \tilde{\beta} ^2\right) y-2 \dot{x} \dot{y}    \right]   \partial_{\dot{y}}  
  	  \notag 
  \end{eqnarray}
  As expected, the generator $X_1$ coincides with the dynamical vector field $V$ and
  represents the trivial symmetry associated with time translations. The second generator $X_2$ provides a genuinely non-trivial Lie point symmetry of the
  H\'enon-Heiles flow. The two symmetries commute and are symmetries of both Hamiltonians 
  \begin{equation}
  [X_1,X_2]=0,  \quad        X_1	\left(  {\cal H}_1 \right) =0,  \quad X_1	\left(  {\cal H}_2 \right) =0, \quad
             X_2	\left(  {\cal H}_1 \right) =0,  \quad X_2	\left(  {\cal H}_2 \right) =0.
  \end{equation}
  This again demonstrates the compatibility between the Lie symmetry structure and the
  conformal bi-Hamiltonian formulation of the integrable H\'enon-Heiles system.
  
  \subsection{ H\'enon-Heiles versus interacting PU picture}
  
  Using the elimination procedure described after (\ref{FFF}), we may express the two
  H\'enon-Heiles Hamiltonians entirely in terms of the reduced variables
  $(x,\dot x,\ddot x,\dddot x)$.
  In this way the first Hamiltonian simply reproduces the energy constraint
  ${\cal H}_1=E$, while the second Hamiltonian reduces, up to an additive constant,
  to the interacting PU Hamiltonian
  \begin{equation}
  	{\cal H}_1 (x,\dot{x},\ddot{x},\dddot{x}) =E, \quad
  	{\cal H}_2 (x,\dot{x},\ddot{x},\dddot{x}) = 2 {\cal H}_{\text{PUI}}(x,\dot{x} , \ddot{x},\dddot{x}) - 2(\tilde{\alpha} - 4 \tilde{\beta}) E.
  \end{equation}   
  We emphasise that, under the H\'enon-Heiles reduction, only the second
  H\'enon-Heiles Hamiltonian gives rise to the interacting PU Hamiltonian,
  whereas the second Hamiltonian of the reduced PU system does not admit a
  counterpart at the level of the original H\'enon-Heiles formulation. Here the elimination of the H\'enon-Heiles variables is understood purely at the level
  of the equations of motion and does not correspond to a Hamiltonian or symplectic
  reduction. This asymmetry shows that the conformal bi-Hamiltonian structure of the interacting PU system is genuinely richer than that inherited from the Hénon-Heiles formulation.

  \section{Classical solutions}
  
   \subsection{Direct numerical analysis}
   Before addressing the analytic construction of the classical solutions, we first
   investigate the interacting PU equation of motion (\ref{equmPUI}) numerically.  This provides a
   direct picture of the dynamics of the fourth-order system and serves as a motivation
   for the subsequent analytic analysis. We integrate the equation (\ref{equmPUI}) for representative choices of the parameters and initial conditions, working directly with the fourth-order equation and without relying on the auxiliary H\'enon-Heiles formulation or on the existence of conserved quantities.
   
\begin{figure}[h]
  	\begin{minipage}[b]{\textwidth}       
  		\includegraphics[width=0.5\textwidth]{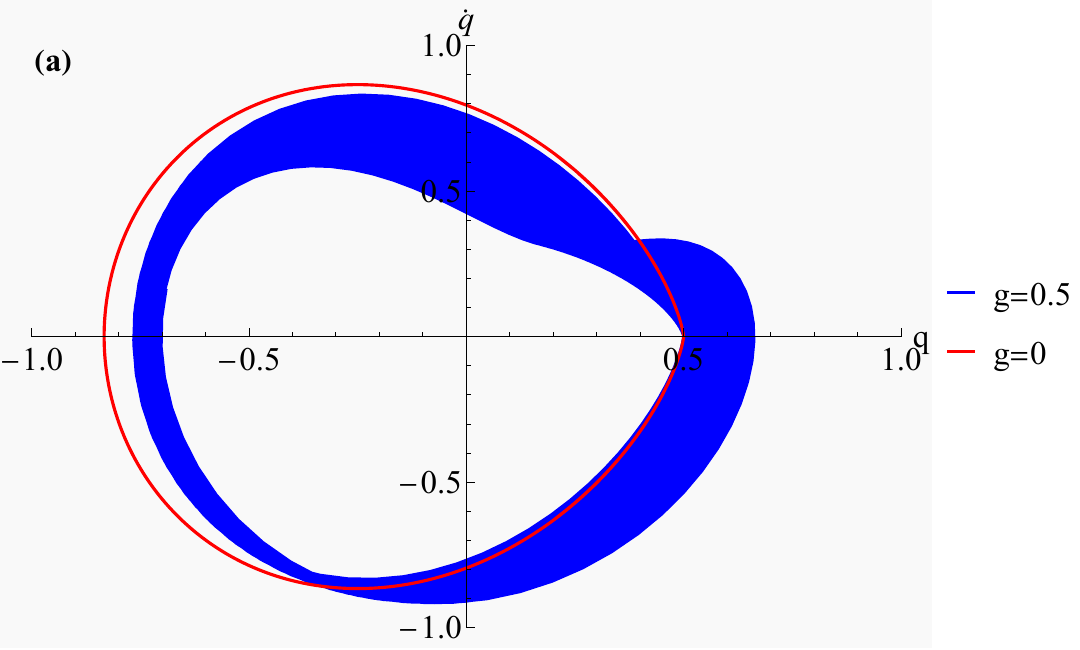}
  		\includegraphics[width=0.5\textwidth]{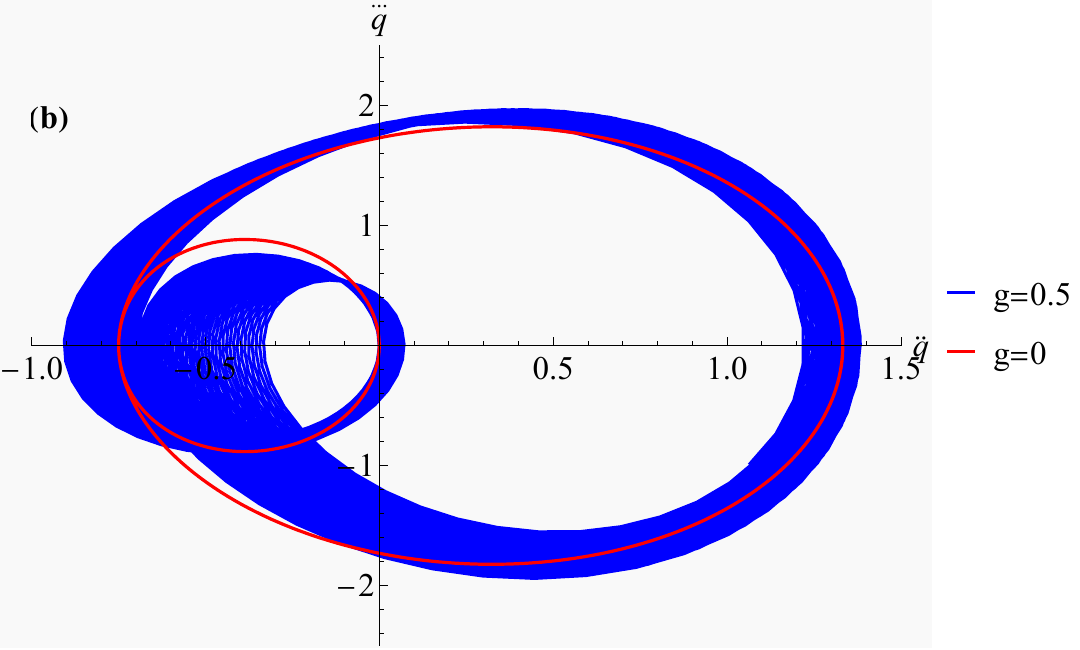}
  		\includegraphics[width=0.5\textwidth]{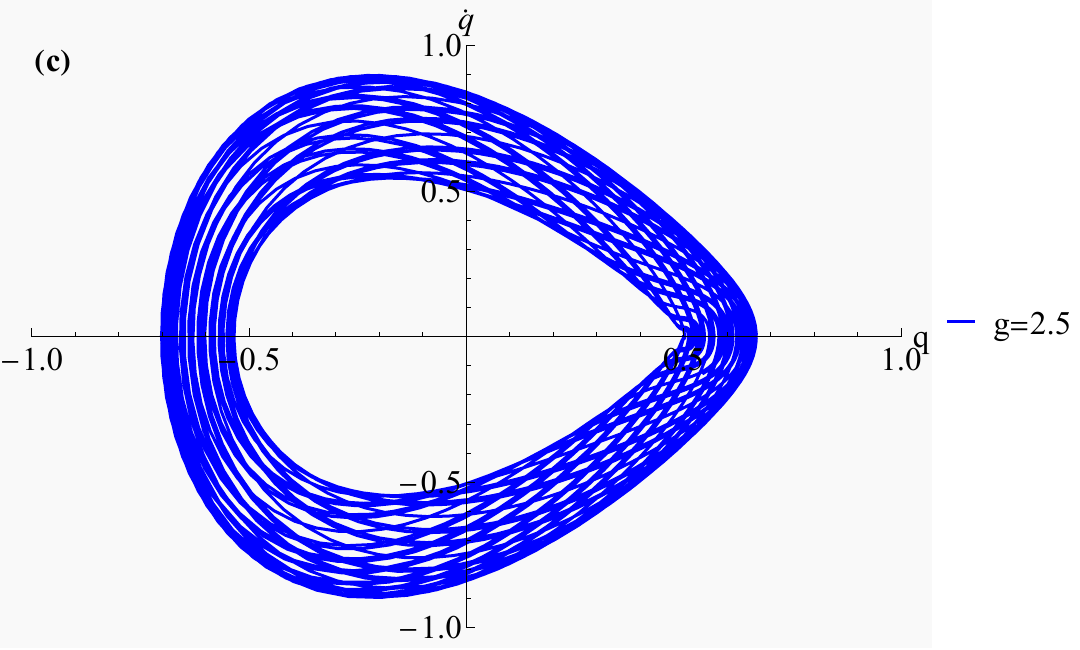}
  		\includegraphics[width=0.5\textwidth]{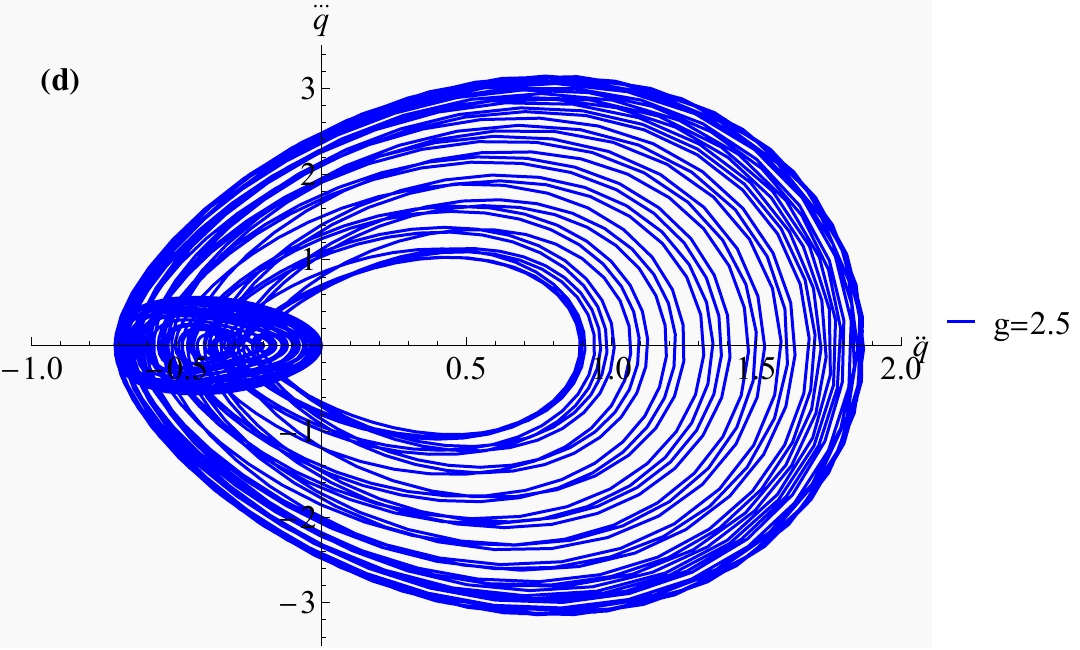}
  		\includegraphics[width=0.5\textwidth]{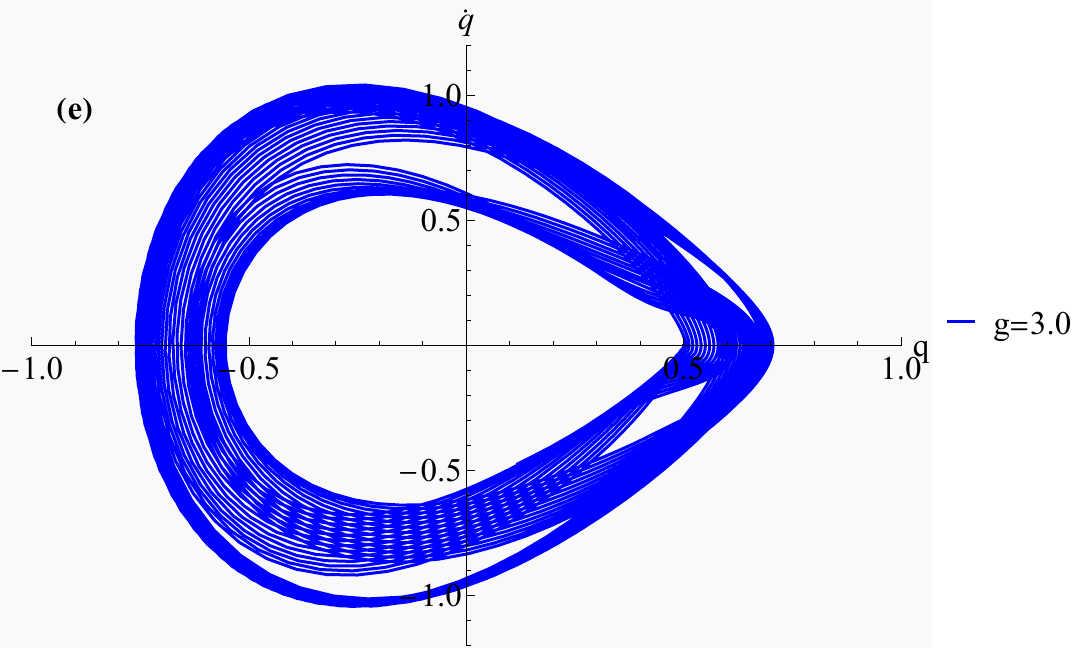}
  		\includegraphics[width=0.5\textwidth]{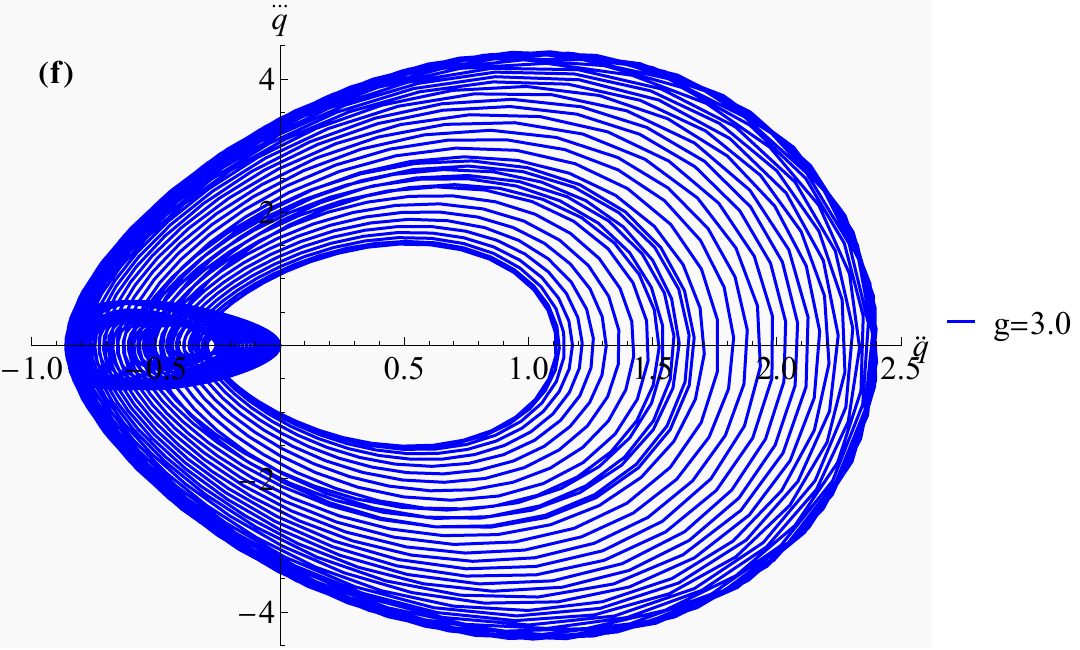}
  		\includegraphics[width=0.5\textwidth]{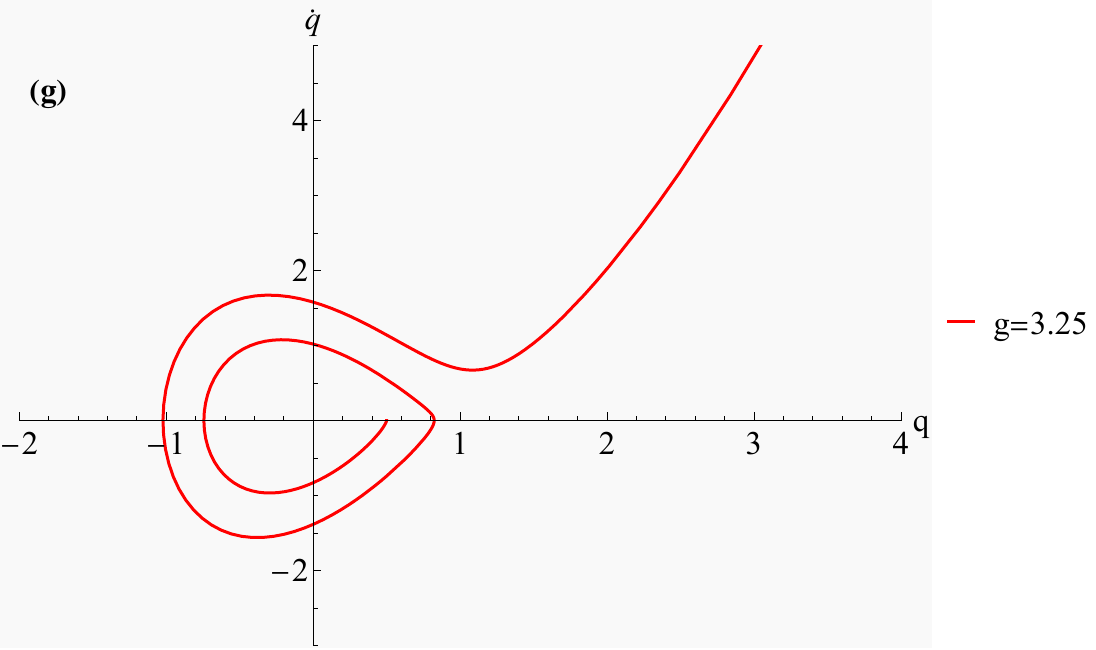}
  		\includegraphics[width=0.5\textwidth]{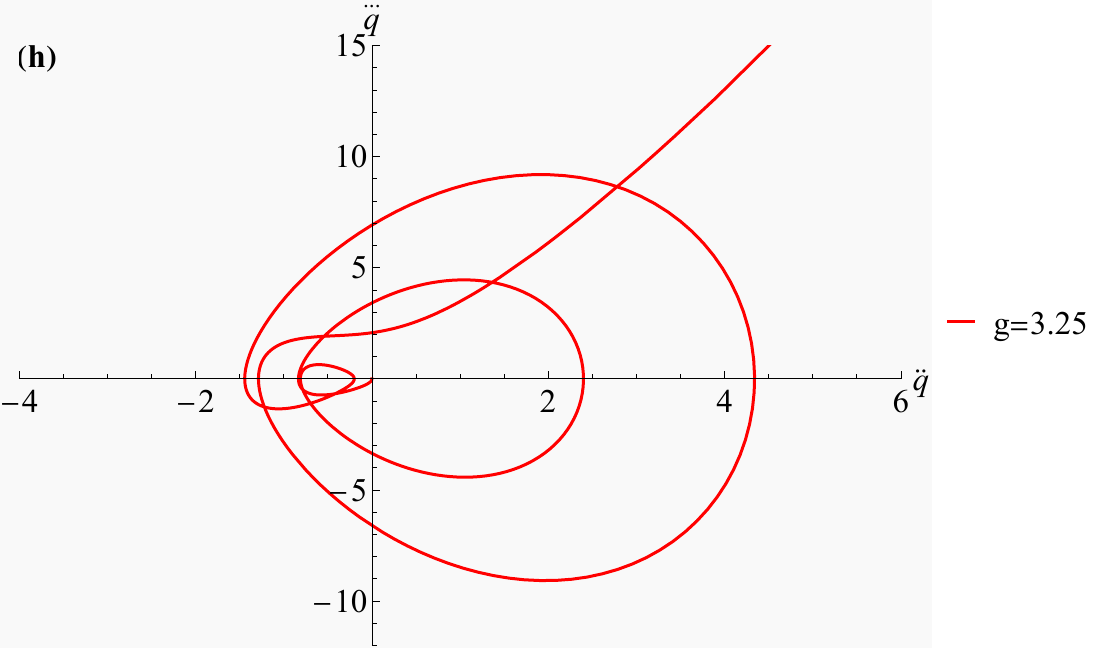}
  	\end{minipage}   
  	\caption{$(q,\dot{q})$ and $(\ddot{q},\dddot{q})$ phase space, for the parameters $\alpha=5$, $\beta=4$, $E=-0.5$, initial values $q(t=0)=0.5$ , $\dot{q}(t=0)  = \ddot{q}(t=0)=\dddot{q}(t=0) =0 $ for different values of the coupling constant $g$. The time is running from $t_0=0$ to $t=250$ in panels (a)-(f) and from $t_0=0$ to $t=60$ in panels (g) and (h).} 
  	\label{phasespace}
  \end{figure}

     Our results for some specific values of the model parameters and initial values are depicted as phase space portraits in figure \ref{phasespace}. The numerical solutions exhibit bounded and regular motion over long integration times.  In particular, the coordinate $q(t)$ displays periodic behaviour and no indication of runaway solutions or secular growth is observed in some parameter regimes. This behaviour is highly nontrivial for an interacting higher-derivative system and  stands in marked contrast to the generic instability expected from
     Ostrogradsky-type dynamics. For larger values of the coupling constant the dynamics undergoes a qualitative change and unbounded trajectories appear, as illustrated in panels (g) and (h) of figure \ref{phasespace}.
  
\begin{figure}[h]
  	      
  	\begin{minipage}[b]{\textwidth}     
  		\centering    
  		\includegraphics[width=0.75\textwidth]{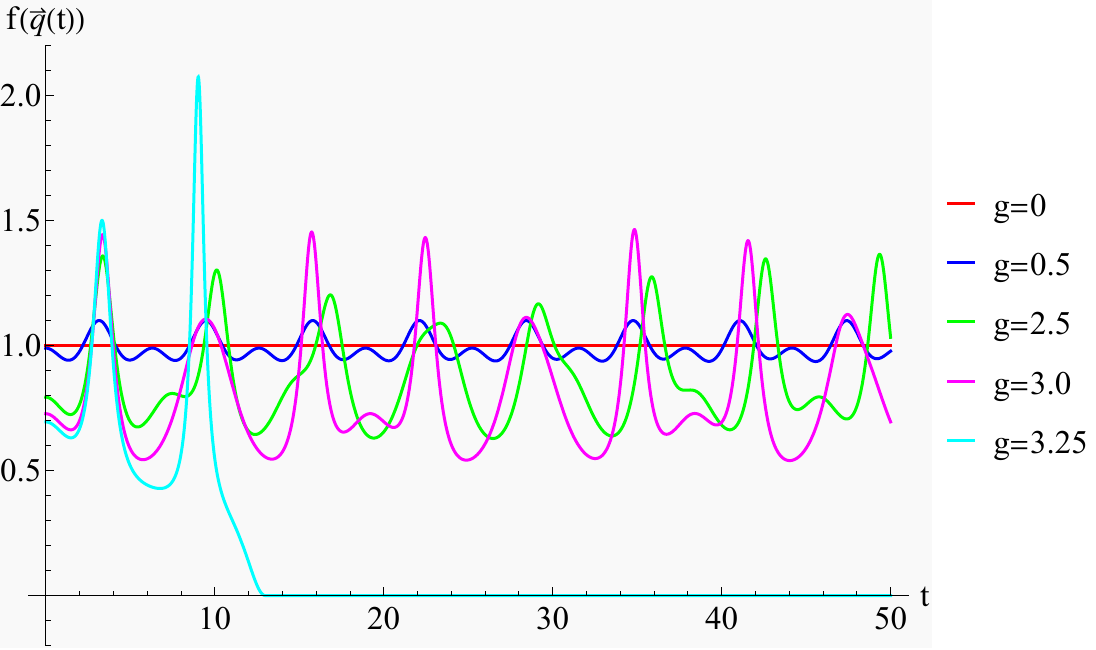}
  	\end{minipage}   
  	\caption{Conformal factor in equation (\ref{confact}) as function of $t$ for the solutions from figure \ref{phasespace}.} 
  	\label{conformalfact}
  \end{figure}

  An important question is whether the conformal factor $f(\vec q)$ in (\ref{confact}) can become ill-defined along the flow. Given the form in (\ref{confact}) the conformal representation breaks down only if the denominator vanishes, that is, on the hypersurface
  $ 3\beta+2g^2 q^2-2g\ddot q=0$.
  To assess this issue for the trajectories considered in figure \ref{phasespace}, we plot the conformal factor $f(\vec q(t))$ as a function of time in figure \ref{conformalfact}. For the bounded solutions the conformal factor remains finite throughout the integration interval. By contrast, for the runaway trajectory $f(\vec q(t))$ rapidly decreases and approaches zero after a critical time, without crossing the singular hypersurface defined by the vanishing of the denominator. For those solutions the time reparametrisation $d\tau=f\,dt$ becomes degenerate and the conformal bi-Hamiltonian description in terms of the parameter $\tau$ is effectively restricted to a finite time interval.
  
  The regular and bounded character of the numerical solutions strongly suggests that
  the interacting PU system considered here possesses additional hidden
  structure.  In the following subsections we show that this behaviour can be explained
  by an explicit reduction to an integrable H\'enon-Heiles system and by a subsequent
  separation of variables.
 
  \subsection{Algebraic reduction}
  In order to explain the regular behaviour observed in the numerical analysis, we now
  construct the classical solutions analytically. Since we have two integrals of motion the system is Liouville integrable. We can utilise the integrals to solve the classical equations of motion. For this purpose we introduce the parametrisation
  \begin{equation}
  	    \dot{x} =: p, \qquad \dot{y} =: q, \qquad r:= \frac{ \dot{y}}{ \dot{x}}= \frac{q}{p} =\frac{dy}{dx} .
  \end{equation}
  Setting now ${\cal H}_1 =E_1$, with constant $E_1$, we solve this equation to
    \begin{equation}
       p^2 = \frac{F(x,y,E_1)}{1+r^2}, \quad \text{with}\,\, F(x,y,E_1) :=  2 E_1+4 x^3+2 x y^2-x^2 \tilde{\alpha} -y^2 \tilde{\beta}  . \label{p2}
  \end{equation}
  Using this relation to replace $p^2$ in the equation ${\cal H}_2 =E_2$, it is converted into a quadratic relation for $r$ 
   \begin{equation}
  	A(x,y,E_1,E_2) r^2 +	B(x,y,E_1) r + 	C(x,y,E_2)  =0,    \label{quadratic}
  \end{equation}
with
\begin{eqnarray}
	A(x,y,E_1,E_2) &=&16 x^4+y^4-2 x y^2 \tilde{\alpha} +x^2 \left[12 y^2+\tilde{\alpha}  (\tilde{\alpha} -4 \tilde{\beta} )\right]-8 x^3 (\tilde{\alpha} -2 \tilde{\beta} )  \qquad  \\ &&+ (8 x-2 \tilde{\alpha} +8 \tilde{\beta} )E_1-E_2 , \notag \\ 
	B(x,y,E_1)  &=&  4 x^2 y \tilde{\alpha} +4 y^3 \tilde{\beta} -16 x^3 y-8 x y^3-8 E_1 y ,\\
	C(x,y,E_2)  &=& 4 x^2 y^2+y^4-4 x y^2 \tilde{\beta} -y^2 \tilde{\alpha}  \tilde{\beta} +4 y^2 \tilde{\beta} ^2-E_2 .
\end{eqnarray}
Solving (\ref{quadratic}) gives the first order ODE
\begin{equation}
	\frac{dy}{dx} = \frac{-B\pm \sqrt{B^2- 4 A C}}{ 2 A} =r.    \label{dydx}
\end{equation}
Integrating this equation yields $y(x)$. From equation (\ref{p2}) we obtain
\begin{equation}
	\frac{dx}{dt} =   \pm \sqrt{ \frac{F[x,y(x), E_1]}{ 1 + r^2[x,y(x),E_1,E_2 ]  }    },
\end{equation}
which integrated yields
\begin{equation}
	t - t_0 = \int_{x_0}^x  \sqrt{ \frac{ 1 + r^2[s,y(s),E_1,E_2 ]  }{F[s,y(s), E_1]}    } ds .   \label{lastint}
\end{equation} 
Thus, solving (\ref{dydx}) numerically, we can use the result to integrate out (\ref{lastint}) obtaining $t(x)$, which upon inversion gives a solution $x(t)$ of the fourth order PU-equation with interaction (\ref{HHequom}).  

\subsection{Separation of variables from conformal bi-Hamiltonians}

In the previous subsections we established a conformal bi-Hamiltonian structure for
the integrable H\'enon-Heiles system. In general, when the Hamiltonian formulation is
non-canonical, as in the interacting PU system discussed above, the
standard geometric mechanism for separation of variables based on Killing tensors is
no longer available, since the dynamics is defined on a Poisson manifold without a
distinguished configuration-space metric. In such a situation, separability is
governed instead by the spectral properties of the conformal recursion operator
$R_c=f^{-1}J_2J_1^{-1}$ acting directly on phase space.

In the present H\'enon-Heiles formulation, however, the Hamiltonian structure is
canonical and the system belongs to the class of natural Hamiltonians. The second
Hamiltonian is quadratic in the momenta and therefore defines a Killing tensor on the
configuration space. In this case the recursion operator $ R=J_2J_1^{-1}$
constructed from the two Poisson tensors admits, when restricted to the configuration
space, the mixed tensor $K^{i}{}_{j}$ associated with the Killing tensor
$K^{ij}$. The fact that the separation coordinates obtained from the Killing tensor
coincide with the non-trivial eigenvalues of the recursion operator thus provides a
direct and non-trivial consistency check of the Poisson tensor $J_2$ constructed in
the previous subsection.

\begin{figure}[h]
 		\begin{minipage}[b]{\textwidth}      
 				\centering
 			\includegraphics[width=0.75\textwidth]{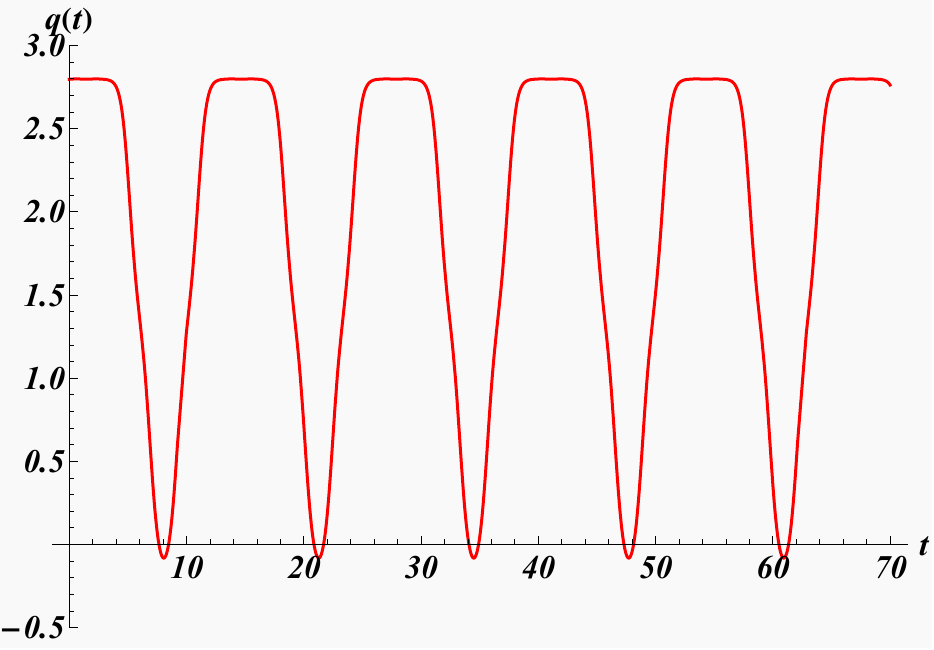}
 		\end{minipage}   
 		\caption{Periodic solution to the fourth order equation (\ref{equmPUI}) for the parameters $\tilde{\alpha} = 4.1$, $\tilde{\beta} = 2.35$, $E=E_1=-0.181609$, $E_2=- 8.15687 $, $a_+ = 1.35991$, $a_- = 5.64009$ and  $a_3=16.6$.} 
 		\label{solution4th}
 	\end{figure}

These observations together with the work in \cite{ravoson1993separability} suggest that our system is likely to be
separable. We therefore follow the St\"ackel algorithm
\cite{stackel1891integration}. According to Benenti's theorem
\cite{benenti1997intrinsic}, if a Hamiltonian system admits orthogonal separation
coordinates, then these coordinates coincide with the eigenvalues of the Killing
tensor $K^{ij}$ defined through
\begin{equation}
	{\cal H}_2 = K^{ij}(x,y)\,p_i p_j + V(x,y).
\end{equation}
In our case we have
\begin{equation}
	K= \left(
	\begin{array}{cc}
		0 & -4 y \\
		-4 y & 2 (4 \tilde{\beta} -\tilde{\alpha} )+8 x
	\end{array}
	\right),
\end{equation}
with eigenvalues
\begin{equation}
	u ,v = 4 x-\tilde{\alpha} +4 \tilde{\beta} \pm
	\sqrt{16 y^2+(\tilde{\alpha} -4 \tilde{\beta} -4 x)^2}.
\end{equation}
As anticipated, these eigenvalues coincide with those of the recursion operator $R=J_2 J_1^{-1}$, apart from a factor 4 due to the fact that we dropped the factor $1/2$ in the kinetic term.
	Inverting these relations we obtain $x(u,v)$ and $y(u,v)$ as
	\begin{equation}
		x = \frac{1}{8} (u+v+2 \tilde{\alpha} -8 \tilde{\beta} ), \qquad  y= \pm \frac{1}{4} i \sqrt{u v } . \label{xyuv}
	\end{equation}
	The appearance of $i=\sqrt{-1}$ reflects the chosen parametrisation of the
	separation variables, but as we see below the resulting physical coordinates $(x,y)$ remain real on the relevant real slice of the separated dynamics. From 
		\begin{equation}
		p_x \left(  \frac{\partial x}{ \partial u} du   + \frac{\partial x}{ \partial v} dv      \right) +
		p_y \left(  \frac{\partial y}{ \partial u} du   + \frac{\partial y}{ \partial v} dv      \right) =
		p_u du + p_v dv,
	\end{equation}
	we find the corresponding transformations for the momenta
	\begin{equation}
		p_x = 8\frac{ p_u u-p_v v }{u-v}, \qquad p_y = \pm 8 i\frac{ p_u-p_v }{u-v}  \sqrt{u v}.
	\end{equation}
	Using these transformations the Hamiltonians transform into the St\"ackel representation
	\begin{eqnarray}
		 {\cal H}_1(u,v,p_u,p_v) &=& \frac{32}{u-v} \left( p_u^2 u - p_v^2 v   \right)   + \frac{F(u)-F(v)}{u-v}, \\
		  {\cal H}_2(u,v,p_u,p_v)&= &\frac{32}{u-v} uv \left( p_u^2  - p_v^2    \right)   + \frac{u v}{u-v}   \left[ \frac{F(u)}{u}-\frac{F(v)}{v}\right], 
	\end{eqnarray}
	where we used the fourth order polynomial auxiliary function
	\begin{equation}
		F(x) = \frac{1}{256} x (8 \tilde{\beta} -x) (x+2 \tilde{\alpha} -8 \tilde{\beta} )^2 .
	\end{equation}
	The separation is obtained by noting that
	\begin{equation}
	    \left(
	    \begin{array}{c}
	    	\Phi \left(u,p_u\right) \\
	    	\Phi \left(v,p_v\right) \\
	    \end{array}
	    \right)  = S \left(
	    \begin{array}{c}
	    	{\cal H}_1 \\
	    {\cal H}_2\\
	    \end{array}
	    \right)   \qquad \text{with} \,\, S=\left(
	    \begin{array}{cc}
	    	u & -1 \\
	    	v & -1 \\
	    \end{array}
	    \right),
	\end{equation}
	where $S$ is the St\"ackel matrix and
		\begin{equation}
	\Phi \left(\lambda, \mu \right) := 32 \lambda \mu^2 + F(\lambda). 
	\end{equation}
	Setting the Hamiltonians to constants, $	{\cal H}_1 =E_1$, $	{\cal H}_2 =E_2$ yields the two separated equations in form of two identical elliptic curves
		\begin{equation}
	            w^2(\lambda, p_\lambda) = P_4(\lambda),\qquad     \lambda = u,v.     \label{ellcurve}
	\end{equation}
	with
		\begin{eqnarray}
			 w&=& \sqrt{8192 \lambda}  p_\lambda  \\
		  P_4&=& \lambda ^4+4 (\tilde{\alpha} -6 \tilde{\beta} ) \lambda ^3+4 (\tilde{\alpha} -12 \tilde{\beta} ) (\tilde{\alpha} -4 \tilde{\beta} ) \lambda ^2+32 \left(8 E_1-(\tilde{\alpha} -4 \tilde{\beta} )^2 \tilde{\beta} \right) \lambda -256 E_2  .\notag
	\end{eqnarray}
	As our aim is to find $u(t)$ and $v(t)$ we use Hamilton's equation to compute
		\begin{equation}
		\frac{du}{dt} = \frac{\partial 	{\cal H}_1 }{\partial p_u } = \frac{64}{u-v} u p_u, \qquad 
			\frac{dv}{dt} = \frac{\partial 	{\cal H}_1 }{\partial p_v } = - \frac{64}{u-v} v p_v.
	\end{equation}
	Extracting $p_u$ and $p_v$ from (\ref{ellcurve}) we obtain
		\begin{equation}
		\frac{du}{dt} = \frac{1}{\sqrt{2}} \frac{1}{u-v}   \sqrt{u P_4(u) }   \qquad 
		\frac{dv}{dt} = -\frac{1}{\sqrt{2}} \frac{1}{u-v}   \sqrt{v P_4(v) } .
	\end{equation}
	Introducing the St\"ackel time $\tau$ with
	\begin{equation}
		d\tau = \frac{dt}{u-v },   \label{Stackeltime}
	\end{equation}
	the equations decouple into 
		\begin{equation}
		\frac{du}{d \tau}  =\frac{1}{\sqrt{2}} \sqrt{u P_4(u) }, \qquad
		\frac{dv}{d \tau}  =-\frac{1}{\sqrt{2}} \sqrt{v P_4(v) }.    \label{uvint}
	\end{equation}
	Separating variables in each equation gives
	\begin{equation}
		\tau - \tau_0  = \sqrt{2} \int_{u_0}^u \frac{1}{\sqrt{u P_4(u)}} du , \qquad
		\tau - \tau_0  = -\sqrt{2} \int_{v_0}^v \frac{1}{\sqrt{v P_4(v)}} dv .\qquad
 	\end{equation}
 	We may solve the integrals in terms of elliptic function by assuming that $P_4(u)$ has a degenerate root. This choice allows for closed-form elliptic integrals rather than general hyperelliptic integrals. We may re-write 
 	\begin{equation}
 		P_4(u)= (u-a_+)(u-a_-)(u-a_3)^2,
 	\end{equation}
 	with
 	\begin{eqnarray}
 		E_1&=& -\frac{1}{64} \left(2 \tilde{\alpha} +a_3-8 \tilde{\beta} \right) \left[-2 \tilde{\beta}  \left(2 \tilde{\alpha} +5 a_3\right)+a_3 \left(\tilde{\alpha} +a_3\right)+16 \tilde{\beta} ^2\right],\\
 		E_2 &=& -\frac{1}{256} a_3^2 \left(2 \tilde{\alpha} +a_3-8 \tilde{\beta} \right) \left(2 \tilde{\alpha} +3 a_3-24 \tilde{\beta} \right),\\
 		a_\pm &=& 12 \tilde{\beta} -2 \tilde{\alpha} -a_3\pm \sqrt{8 \tilde{\beta}  \left(2 \tilde{\alpha} +3 a_3\right)-2 a_3 \left(2 \tilde{\alpha} +a_3\right)-48 \tilde{\beta} ^2} .
 	\end{eqnarray}	
 	In the regime where $0<a_- < a_+  < u$ we find
 	\begin{equation}
 	\tau(u)= \frac{2 \sqrt{2}}{a_3 \sqrt{a_-}}   \left\{  F\left[ \arcsin\left( \sqrt{\frac{a_-}{u }} \right)     \right. \left\vert \frac{a_+}{a_-}        \right]  - \Pi\left[ \frac{a_3}{a_-} ;    \arcsin\left( \sqrt{\frac{a_-}{u }} \right)     \right. \left\vert \frac{a_+}{a_-}        \right]        \right\}  + \tau(u_0),
 	\label{ellipticFP}
 	\end{equation}
 	where $F(\phi\vert m)$ is the elliptic integral of the first kind and $\Pi(n; \phi\vert m)$ is an incomplete elliptic integral of the third kind. From (\ref{uvint}) it follows that $\tau(v) = -[\tau(u=v)-\tau(u_0)] + \tau(v_0) $.
 	We cannot invert the expression in (\ref{ellipticFP}) in terms of elementary functions to get $u(\tau)$ and $v(\tau)$, so at this point we shall resort to a numerical evaluation. It is clear from the multi valuedness of the $\arcsin$-function that $u(\tau)$ will be a periodic function. Having obtained these two functions, we convert the St\"ackel time to ordinary time by means of (\ref{Stackeltime})
 	\begin{equation}
 	t-t_0 = \int_{\tau_0}^{\tau} \left[ u(s)-v(s) \right] ds,
 	\end{equation}
 	which gives us $t(\tau)$. Again we invert this function numerically in order to obtain $\tau(t)$.	Finally, using (\ref{xyuv}) we reconstruct
 	\begin{equation}
 		x(t) = \frac{1}{8} \left\{ u\left[ \tau(t)    \right] +v\left[ \tau(t)    \right] +2 \tilde{\alpha} -8 \tilde{\beta} \right\}.
 	\end{equation}
 	For the special choice of couplings $g_1=-1$, $g_2=6$, and upon identifying $x(t)=q(t)$,
 	this function provides, through the H\'enon-Heiles reduction discussed in
 	section 3, a solution of the interacting PU equation
 	(\ref{equmPUI}). For a specific choice of parameters we depict the solution in figure \ref{solution4th} demonstrating that there are indeed periodic solutions to this equation.
 	
 	\section{Conclusions}
 	We have presented a concrete and explicit example of an interacting higher-time-derivative system whose classical dynamics remains analytically tractable. For a PU oscillator supplemented with a Landau-Ginzburg type interaction we have shown that the resulting fourth-order equation of motion admits a conformal bi-Hamiltonian formulation and possesses a non-trivial Lie symmetry in addition to the time-translation symmetry. These structures provide a geometric framework for the interacting higher-derivative dynamics similar to that of well-known finite-dimensional integrable systems.
 	
 	By establishing an explicit elimination procedure relating the interacting PU equation to a special integrable generalised Hénon–Heiles system, we demonstrated that the higher-derivative model inherits the integrability properties of the latter. In particular, the correspondence allows one to identify a second conserved quantity, to clarify the geometric origin of separability and to construct separation variables in a systematic manner. This connection also explains why the interacting PU dynamics admits periodic classical solutions despite the presence of higher time derivatives and interaction terms.
 	
 	Complementing the analytic construction, we performed a direct numerical study of the fourth-order PU equation itself. The numerical phase-space portraits show that, for suitable choices of parameters and initial conditions, the interacting higher-derivative dynamics exhibits bounded and regular motion over long integration times. This behaviour is highly non-trivial in view of the generic instabilities associated with Ostrogradsky-type systems and provides an independent confirmation of the integrable structure uncovered in the analytic treatment. At the same time, the numerical results indicate that this regular behaviour is not universal throughout the entire parameter range and that qualitative changes of the dynamics occur when the interaction strength is increased.
 	
 	The conformal character of the bi-Hamiltonian structure is governed by a non-constant conformal factor. 
 	The corresponding singular hypersurface in phase space, defined by the vanishing of the denominator of the conformal factor, represents a genuine geometric obstruction for the conformal Hamiltonian representation. 
 	The families of periodic solutions constructed in this work remain away from this singular set, and for these trajectories the conformal factor stays finite along the flow, so that the conformal bi-Hamiltonian formulation and the associated separation procedure are well defined. 
 	By contrast, for the runaway solutions observed in the numerical analysis the conformal factor rapidly decreases and approaches zero after a finite time, so that the associated time reparametrisation becomes degenerate and the conformal Hamiltonian description is effectively restricted to a finite time interval. 
 	A more detailed analysis of the dynamical role of the singular hypersurface, and of its relation to the transition between bounded and unbounded motion, constitutes an interesting direction for future investigations.
 	
 	Several natural extensions of the present work are suggestive. On the classical side, it would be of interest to explore whether further classes of interaction terms can be incorporated while preserving a conformal bi-Hamiltonian structure or an explicit link to known integrable systems. On the quantum side, the existence of a geometric and integrable structure for an interacting higher-derivative model raises the question of whether ghost-free quantisation schemes developed for the free PU oscillator can be meaningfully extended to this interacting setting. We expect that the explicit and fully controlled example presented here provides a useful starting point for addressing these problems.
 	
 	 \medskip
 	
 	\noindent {\bf Acknowledgments}: AlF acknowledges funding from the Max Planck Society’s Lise Meitner Excellence Program 2.0. 
 	
\newif\ifabfull\abfulltrue


\begin{thebibliography}{10}
	
	\bibitem{pais1950field}
	A.~Pais and G.~E. Uhlenbeck,
	\newblock On field theories with non-localized action,
	\newblock Phys. Rev. {\bf 79}(1), 145 (1950).
	
	\bibitem{ghostconst}
	T.-J. Chen, M.~Fasiello, E.~A. Lim, and A.~J. Tolley,
	\newblock Higher derivative theories with constraints: Exorcising
	Ostrogradski's Ghost,
	\newblock J. Cos. Astro. Phys. {\bf 2013}(02), 042 (2013).
	
	\bibitem{motohashi1}
	H.~Motohashi, K.~Noui, T.~Suyama, M.~Yamaguchi, and D.~Langlois,
	\newblock Healthy degenerate theories with higher derivatives,
	\newblock J. Cosmol. Astropart. Phys. {\bf 2016}(07), 033 (2016).
	
	\bibitem{motohashi4}
	H.~Motohashi and T.~Suyama,
	\newblock Quantum Ostrogradsky theorem,
	\newblock JHEP {\bf 2020}(9), 1--10 (2020).
	
	\bibitem{bolonek2005ham}
	K.~Bolonek and P.~Kosinski,
	\newblock Hamiltonian Structures for Pais--Uhlenbeck Oscillator,
	\newblock Acta Phys. Polon. B {\bf 36}(6), 2115 (2005).
	
	\bibitem{bender2008no}
	C.~M. Bender and P.~D. Mannheim,
	\newblock No-ghost theorem for the fourth-order derivative Pais-Uhlenbeck
	oscillator model,
	\newblock Phys. Rev. Lett. {\bf 100}(11), 110402 (2008).
	
	\bibitem{smilga2009comments}
	A.~Smilga,
	\newblock Comments on the dynamics of the Pais-Uhlenbeck oscillator,
	\newblock SIGMA {\bf 5}, 017 (2009).
	
	\bibitem{most2010h}
	A.~Mostafazadeh,
	\newblock A Hamiltonian formulation of the Pais--Uhlenbeck oscillator that
	yields a stable and unitary quantum system,
	\newblock Phys. Lett. A {\bf 375}(2), 93--98 (2010).
	
	\bibitem{mosta2011im}
	A.~Mostafazadeh,
	\newblock Imaginary-scaling versus indefinite-metric quantization of the
	Pais-Uhlenbeck oscillator,
	\newblock Phys. Rev. D {\bf 84}(10), 105018 (2011).
	
	\bibitem{andrzejewski2014ham}
	K.~Andrzejewski,
	\newblock Hamiltonian formalisms and symmetries of the Pais--Uhlenbeck
	oscillator,
	\newblock Nuclear Physics B {\bf 889}, 333--350 (2014).
	
	\bibitem{Sugg3}
	M.~Avendano-Camacho, J.~A. Vallejo, and Y.~Vorobiev,
	\newblock A perturbation theory approach to the stability of the Pais-Uhlenbeck
	oscillator,
	\newblock J. Math. Phys. {\bf 58}(9) (2017).
	
	\bibitem{smilga2017class}
	A.~Smilga,
	\newblock Classical and quantum dynamics of higher-derivative systems,
	\newblock Int. J. Mod. Phys. A {\bf 32}(33), 1730025 (2017).
	
	\bibitem{fring2025quant}
	A.~Fring, T.~Taira, and B.~Turner,
	\newblock Quantisations of exactly solvable ghostly models,
	\newblock J. Phys. A: Math. Theor. {\bf 58}(23), 235301 (2025).
	
	\bibitem{FFT}
	A.~Felski, A.~Fring, and B.~Turner,
	\newblock Lie symmetries and Ghost-free representations of the Pais--Uhlenbeck
	model,
	\newblock Mod. Phys. Lett. A , 2650019 (2026).
	
	\bibitem{felski2026three}
	A.~Felski, A.~Fring, and B.~Turner,
	\newblock Three-dimensional ghost-free representations of the Pais-Uhlenbeck
	model from Tri-Hamiltonians,
	\newblock Phys. Lett. A , 131332 (2026).
	
	\bibitem{fring2025ghost}
	A.~Fring, T.~Taira, and B.~Turner,
	\newblock Ghost-Free Quantisation of Higher Time-Derivative Theories via
	Non-Unitary Similarity Transformations,
	\newblock arXiv preprint arXiv:2506.21400  (2025).
	
	\bibitem{fring2026spectrum}
	A.~Fring, I.~Marquette, and T.~Taira,
	\newblock Spectrum-generating algebra and intertwiners of the resonant
	Pais-Uhlenbeck oscillator,
	\newblock arXiv preprint arXiv:2601.20752  (2026).
	
	\bibitem{Sugg1}
	M.~Pav{\v{s}}i{\v{c}},
	\newblock Stable self-interacting Pais--Uhlenbeck oscillator,
	\newblock Mod. Phys. Lett. A {\bf 28}(36), 1350165 (2013).
	
	\bibitem{pavvsivc2013pais}
	M.~Pav{\v{s}}i{\v{c}},
	\newblock Pais-Uhlenbeck oscillator with a benign friction force,
	\newblock Phys. Rev. D {\bf 87}(10), 107502 (2013).
	
	\bibitem{pavvsivc2016pais}
	M.~Pav{\v{s}}i{\v{c}},
	\newblock Pais-Uhlenbeck oscillator and negative energies,
	\newblock Int. J. Geo. Meth. Mod. Phys. {\bf 13}(09), 1630015 (2016).
	
	\bibitem{ostrogradsky1850memoire}
	M.~Ostrogradsky,
	\newblock {\em M{\'e}moire sur les {\'e}quations diff{\'e}rentielles relatives
		an probl{\'e}me des isop{\'e}rim{\'e}tres}, volume VI 4,
	\newblock 1850.
	
	\bibitem{morosi1997quasi}
	C.~Morosi and G.~Tondo,
	\newblock Quasi-bi-Hamiltonian systems and separability,
	\newblock J. Phys. A: Math. Gen. {\bf 30}(8), 2799 (1997).
	
	\bibitem{crampin2002bi}
	M.~Crampin and W.~Sarlet,
	\newblock Bi-quasi-Hamiltonian systems,
	\newblock J. Math. Phys. {\bf 43}(5), 2505--2517 (2002).
	
	\bibitem{falqui2003separation}
	G.~Falqui and M.~Pedroni,
	\newblock Separation of variables for bi-Hamiltonian systems,
	\newblock Math. Phys., Anal. Geo. {\bf 6}(2), 139--179 (2003).
	
	\bibitem{henonheils}
	M.~H{\'e}non and C.~Heiles,
	\newblock The applicability of the third integral of motion: some numerical
	experiments,
	\newblock Astro. J. 69 {\bf 69}, 73 (1964).
	
	\bibitem{fordy1991henon}
	A.~P. Fordy,
	\newblock The H{\'e}non-Heiles system revisited,
	\newblock Physica D {\bf 52}, 204--210 (1991).
	
	\bibitem{caboz1991bi}
	R.~Caboz, V.~Ravoson, and L.~Gavrilov,
	\newblock Bi-Hamiltonian structure of an integrable H{\'e}non-Heiles system,
	\newblock J. Phys. A: Math. Gen. {\bf 24}(10), L523 (1991).
	
	\bibitem{hydonsymm}
	P.~E. Hydon,
	\newblock {\em Symmetry methods for differential equations: a beginner's
		guide},
	\newblock Number~22, Cambridge University Press, 2000.
	
	\bibitem{ravoson1993separability}
	V.~Ravoson, L.~Gavrilov, and R.~Caboz,
	\newblock Separability and Lax pairs for H{\'e}non--Heiles system,
	\newblock J. Math. Phys. {\bf 34}(6), 2385--2393 (1993).
	
	\bibitem{stackel1891integration}
	P.~St{\"a}ckel,
	\newblock {\"U}ber die Integration der
	Hamilton-Jacobischen-Differentialgleichung mittels der Separation der
	Variabeln,
	\newblock Halle, Phil. Fak., Habil.-Schr.  (1891).
	
	\bibitem{benenti1997intrinsic}
	S.~Benenti,
	\newblock Intrinsic characterization of the variable separation in the
	Hamilton-Jacobi equation,
	\newblock J. Math. Phys. {\bf 38}(12), 6578--6602 (1997).
	
\end{thebibliography}

\end{document}